\documentclass[12pt]{article}
\usepackage{amssymb}
\usepackage{graphicx}    
\usepackage{graphics}
\usepackage{cite}
\usepackage{ulem}
\usepackage{amsmath}
\usepackage{mathtools}

\def\e{{\rm e}}

\def\d{\partial}
\def\l{\left(}                                    
\def\r{\right)}

\newcommand{\be}{\begin{equation}}
\newcommand{\ee}{\end{equation}}
\newcommand{\ba}{\begin{align}}
\newcommand{\ea}{\end{align}}
\newcommand{\bg}{\begin{gather}}
\newcommand{\eg}{\end{gather}}
\newcommand{\bseq}{\begin{subequations}}
\newcommand{\eseq}{\end{subequations}}

\renewcommand{\tanh}{\mathop{\rm th}\nolimits}

\begin{document}
\title{A minimal scale invariant axion solution to the strong CP-problem.}

\author{Anna Tokareva$^{1,2}$\\
\mbox{}$^{1}${\small Institute for Nuclear Research of Russian Academy of
  Sciences, 117312 Moscow,
  Russia}\\  
\mbox{}$^{2}${\small Ecole Polytechnique Fб╢edб╢erale de Lausanne, CH-1015, Lausanne, Switzerland}\\    
}


\maketitle

\begin{abstract} 
We present a scale invariant extension of the Standard model allowing for the Kim-Shifman-Vainstein-Zakharov (KSVZ) axion solution of the strong CP problem in QCD. We add the minimal number of new particles and show that the Peccei-Quinn scalar might be identified with the complex dilaton field. Scale invariance, together with the Peccei-Quinn symmetry, is broken spontaneously near the Planck scale before inflation, which is driven by the Standard Model Higgs field. We present a set of general conditions which makes this scenario viable and an explicit example of an effective theory possessing spontaneous breaking of scale invariance. We show that this description works both for inflation and low-energy physics in the electroweak vacuum. This scenario can provide a self-consistent inflationary stage and, at the same time, successfully avoid the cosmological bounds on the axion. Our general predictions are the existence of coloured TeV mass fermion and the QCD axion. The latter has all the properties of the KSVZ axion but does not contribute to dark matter. This axion can be searched via it's mixing to a photon in an external magnetic field. 
\end{abstract}

\section{Introduction}

Standard Model (SM), being in perfect agreement with many experiments, nevertheless requires an extension which would be able to resolve several problems remaining in cosmology and particle physics, such as neutrino masses, baryon asymmetry, dark matter and dark energy. A huge number of viable models based on different principles (e.g. supersymmetry, grand unification, extra dimensions, etc.), describing everything ranging from the early Universe to the origin of neutrino masses, can be constructed. Therefore, one needs a basic principle to choose the reasonable models. To start with, we can consider the {\it minimal} models allowing to explain all known phenomena by means of the smallest number of extra degrees of freedom which are absent in the SM. A particular model based on the idea of minimality is presented in \cite{newMSM}. This model allows for resolving all well established cosmological and particle physics problems (such as the dark matter and origin of neutrino masses) by means of adding two fermions and two real scalars. Another model, exploiting the principle of the minimal extension of the SM, is the $\nu$MSM (neutrino Minimal Standard Model) \cite{Asaka:2005an,Asaka:2005pn,Shaposhnikov:2006xi}.
It can provide a viable description of primordial inflation, baryogenesis, dark matter, neutrino masses and oscillations. The only extra ingredients of $\nu$MSM are three Majorana neutrinos and a non-minimal coupling between the Higgs field and gravity. In order to explain simultaneously the absence of CP violation in the QCD sector, this model can be further extended with an axion field \cite{Peccei:1977hh}. An example of a complete model accounting for the strong CP problem, dark matter, inflation, neutrino masses and baryon asymmetry was presented in \cite{Salvio:2015cja}. Another model, dubbed SMASH (Standard Model - Axion - Seesaw - Higgs portal inflation), providing with KSVZ mechanism of CP restoration \cite{Kim:1979if,Shifman:1979if}, was proposed recently in \cite{smash}. This model can also provide with a viable cosmology and particle physics by means of extending the SM with an extra complex scalar field, three sterile neutrinos and an extra quark charged under the QCD group.

The next guiding principle in extending the Standard Model, besides minimality, may be the presence of new symmetries. It might help to solve problems arising at the quantum level, as well as some naturalness and hierarchy problems. For example, the smallness of the Higgs boson mass looks to be problematic without any underlying symmetry because, in general, the Higgs mass is affected by the presence of new physics (e.g. massive particles) at high energies \cite{Susskind:1978ms}. This problem might be weakened if the spontaneously broken scale invariance is present \cite{bardeen}. Both dimensional parameters, the Higgs vacuum expectation value (vev) and the Planck mass, may be given by the large vev of the new scalar field called dilaton \cite{Shaposhnikov:2008xb}. Thus, all mass scales are actually provided by one source. Technically, the scale invariance may be conserved at the quantum level, given a special choice for the regularization and renormalization procedure: according to the simplest one, the renormalization scale is replaced by the dilaton field \cite{Englert:1976ep, Wetterich:1987fm, Shaposhnikov:2008xi}. Within this framework, one can obtain a solution to the Standard Model hierarchy problem which is stable against quantum corrections, if there is no new physics between Higgs vev and Planck scale \cite{Vissani:1997ys,Shaposhnikov:2007nj,strumia}. However, the renormalizability of the theory is lost. This should not be thought as a serious drawback since the theory is non-renormalizable in the presence of gravity. Nevertheless, the model may be considered as a valid effective field theory with a cutoff scale \footnote{Hereafter we define the cutoff scale as the lowest energy of the scattering particles at which the tree-level unitarity gets violated. Note that this scale may depend on the background values of the fields \cite{Bezrukov:2010jz}.} close to the Planck scale \cite{Shaposhnikov:2009nk,Bezrukov:2012hx}.

The SMASH model mentioned above includes an explicit intermediate mass scale (the axion decay constant). That provides with a sufficient level of fine tuning needed to obtain both the small Higgs mass and viable inflation (see the discussion in \cite{smash}). In this paper, we address a question whether it is possible to solve this hierarchy problem exploiting the idea of spontaneously broken scale invariance in the absence of heavy particles. In order to make SMASH scale invariant, one needs to add the dilaton field. However, the complex scalar field, responsible for the Peccei-Quinn symmetry breaking, is already present in the model. The novel idea of this paper is to construct a model in which the Peccei-Quinn scalar also plays the role of the dilaton. In this setup, the QCD axion corresponds to a phase of the complex dilaton field. \footnote{The idea that the Peccei-Quinn scalar works as a dilaton providing masses for sterile neutrinos was presented in \cite{Dias:2006th}. However, in this paper, the scale symmetry is broken by the anomaly. In our work, we assume the existence of exact scale symmetry.} Thus, following the idea of minimality, we do not add extra degrees of freedom, as compared to SMASH. At the same time, the scale invariance of the model might help to resolve the hierarchy problem. The purpose of this paper is to construct a model, stable with respect to the quantum corrections, which solves the strong CP problem and provides realistic cosmology and particle physics.
 
In the proposed framework, the scale symmetry must be broken together with the Peccei-Quinn symmetry. It had happened in the early Universe before inflation at the energies around the Planck scale. This means that the Peccei-Quinn symmetry has never been restored thermally. In order to avoid overproduction of the axion dark matter without fine-tuning the initial $\theta$-angle, we require that during inflation QCD was in the strongly coupled phase. A way to obtain that is (following the idea of Ref. \cite{Dvali:1995ce}\footnote{ The string motivated scenarios based on the idea of stronger QCD during inflation can be found in \cite{Banks:1996ea}.}) to let the QCD coupling constant to depend on the values of Higgs and dilaton fields. This would allow for normal QCD phenomenology in the electroweak vacuum, while at large values of the Higgs field (corresponding to inflation), the effective QCD coupling may well be larger than unity. In this regime, the axion can acquire a large mass. Since the fluctuations of the heavy (compared to the Hubble scale) axion during inflation are negligible, it can be left in the minimum of its potential. Therefore, axion dark matter is not produced in this case, avoiding this way all cosmological constraints. 

Let us briefly describe the cosmological scenario, corresponding to the models under consideration. The scale invariance and Peccei-Quinn symmetry were broken by the dilaton vev before inflation driven by the Higgs field\footnote{ In this work, we assume for simplicity that the SM Higgs potential is positive at large field values and has only one minimum. This possibility do not contradict yet the recent data on the Higgs and top quark masses \cite{Degrassi:2012ry}.}. During inflation, QCD was strongly coupled and the axion was heavy, dynamically providing with zero $\theta$-angle. Just after inflation, QCD becomes weakly coupled, with the same constant as it should be in the SM. The axion looses its mass and remains massless until the QCD phase transition. After reheating, the Universe evolves similarly to the $\nu$MSM scenario \cite{Asaka:2005an,Asaka:2005pn,Shaposhnikov:2006xi}. Namely, dark matter is provided by one of the sterile neutrinos. Two remaining sterile neutrinos generate the lepton asymmetry (which transfers to the baryon asymmetry) and SM neutrino masses. In this paper, we discuss several conditions which allow to obtain this scenario and satisfy all cosmological and laboratory constraints.

The paper is organized as follows. In Section 2 we discuss the problems arising in the simplest model for the Higgs inflation in the presence of a complex dilaton field. Then we consider the general scale invariant lagrangian with the complex dilaton being simultaneously the Peccei-Quinn scalar. In Section 3 we discuss several constraints on the action to fulfil all the phenomenological requirements consistently. We give an explicit example of a viable model which is stable against quantum corrections during the inflationary stage. In Section 4 we consider the general predictions for the scale invariant axion scenarios pointing out the distinguishing features for the experiments and observations. Section 5 is devoted to their distinctive observational signatures.

\section{Axion as a phase of dilaton}

The main idea behind this paper is to identify the complex Peccei-Quinn scalar with the dilaton field. In order to construct a model which is stable with respect to quantum corrections and, at the same time, can provide viable cosmology, solution for the strong CP problem and SM physics at low energy, several conditions has to be satisfied. First, QCD axion has to be massive during inflation and its mass must be larger than the Hubble scale. As it was mentioned, this can be reached if the effective QCD coupling is large in the early Universe. Second, the inflaton potential must provide the spectrum of scalar perturbations which is in agreement with the recent Planck data \cite{planck}. The next requirement concerns the extra fermion charged under $SU(3)_c$. It must be heavy during inflation, in order not to suppress the axion mass \cite{diCortona:2015ldu}, while in the SM vacuum its mass should be less than $10$ TeV, to avoid large contributions to the Higgs mass \cite{strumia}. Finally, since we want the model to be a predictive effective theory both during inflation and for low energy physics, we require that the (background-dependent) cutoff scale always must be much larger than all relevant scales. In this section, we discuss the simplest extension of the Higgs dilaton model providing with a solution for the strong CP problem which is viable from the point of view of the low energy physics. However, this model cannot describe the inflationary stage. Therefore,  we consider a general scale invariant lagrangian which may also allow for the viable cosmology.

\subsection{The simplest model}

{\it Scale invariant model with the QCD axion.}
Let us start with the simplest model allowing for the simultaneous breaking of scale and Peccei-Quinn symmetry by the vev of the complex dilaton field charged under the Peccei-Quinn symmetry. We propose the scalar and gravitational sectors of the model to be of the same form as in the Higgs-dilaton model \cite{GarciaBellido:2011de}, except that the dilaton $X$ is a complex field. Thus, the lagrangian looks
\be 
\label{L_scalar_toy}
L_{scalar}=\frac{1}{2}\l\xi |X|^2 +\xi_h h^2\r R - \frac{1}{2}|\d_{\mu} X|^2-\frac{1}{2}(\d_{\mu} h)^2-\frac{\lambda}{4}(h^2-\alpha^2|X|^2)^2-\beta |X|^4,
\ee
where $h$ stands for the radial component of the SM Higgs field in the unitary gauge and the Higgs vacuum expectation value (vev) is provided by the dilaton vev: $\alpha \langle|X|\rangle=v$. $\xi,~\xi_h$ are dimensionless couplings of the fields to gravity. The term $\beta |X|^4$ actually provides a cosmological constant \cite{Shaposhnikov:2008xb}. Therefore, $\beta$ must be tiny, so we omit this term hereafter, since, in this paper, we do not discuss the dark energy problem. To implement the KSVZ axion scenario with the Peccei-Quinn scalar $X$, one has to add two extra Weyl fermions $Q,~\tilde{Q}$ in the fundamental and antifundamental representations of $SU(3)_c$ and with the SM hypercharges $-1/3,~1/3$, respectively. Their Yukawa couplings to the field $X$ are
\be
\label{L_Y}
L_{Y}=y X \tilde{Q} Q + h.c.
\ee 
The presence of the hypercharge for $Q$ allows for their mixing with the SM quarks, making it possible for them to decay, avoiding the problem of their overabundance \cite{Nardi:1990ku,Berezhiani:1992rk}. Furthermore, one needs an additional $U(1)_{PQ}$ symmetry which acts as a chiral rotation on $Q$ and as a phase rotation of $X$. 
In this setup, the phase of the field $X$ would be connected to the axion field,
\be 
X=|X| e^{i a/F_a}, 
\ee
where $F_a$ is defined by the canonical normalization of the axion field. The field $a$ acquires mass due to non-perturbative QCD effects and thus provides with a dynamical relaxation of the QCD $\theta$-angle to zero value. 

This model is a viable solution to the strong CP problem from the point of view of the particle physics. It provides with the invisible axion which has suppressed couplings to all SM particles. Now we consider this particle in a cosmological context and discuss several problems arising in this way.

{\it Cosmology of the axion.}

The scalar sector of the described model coincides with the Higgs-dilaton model \cite{GarciaBellido:2011de}. The difference is that the dilaton field is substituted by the modulus of the Peccei-Quinn field $X$. Thus, the inflationary stage can be obtained in the same way as in the original model. The values of $\xi$ and $\xi_h$ are constrained by the tilt and amplitude of scalar perturbations, $\xi<0.01$, $\xi_h\sim 10^3$. The phase of field $X$ (axion) do not affect the inflationary dynamics, however, it is be displaced from the QCD minimum. This misalignment can lead to the overproduction of the axion dark matter during QCD phase transition.

The standard cosmological scenario for the QCD axion dark matter involves the thermal restoration of the PQ-symmetry after inflation. This can not be done in our case because it would mean the restoration of scale invariance as well. However, the discussed scale invariant models can be the consistent effective description only in the broken phase, because they are defined only for large values of the dilaton field. Therefore, we deal with the scenarios without restoration of the PQ symmetry. The common problem of such models is that, in order to obtain the correct amount of axion dark matter, one needs to tune the initial value of the QCD $\theta$-angle to be at most $10^{-3}$ \cite{Lyth:1989pb}, though its natural value is of order unity. Another problem arising in large scale inflation is an amplitude of axion isocurvature perturbations which is incompatible with the Planck data \cite{Beltran:2006sq}. To avoid both problems, we propose here that during inflation the axion was massive. This can be achieved if the QCD sector is modified during inflation in such a way that the effective QCD coupling constant is large at this stage. We write the non-minimal coupling between the Higgs, dilaton and gluons in the form,
\be 
\label{QCD}
L_{QCD}= -\frac{1}{4 g_0^2} f\l\frac{h}{|X|}\r G_{\mu\nu}G^{\mu\nu}.
\ee
Here the scale invariant function $f$ is such that for large Higgs field values during inflation it makes the coupling constant to be large. At the same time, we require $f(0)=1$, such that we end up with 'standard' QCD in the electroweak vacuum. 

In the strong coupling regime, the PQ symmetry (which is the axion shift symmetry) is broken nonperturbatively leading to the appearance of the axion potential. In the SM vacuum (at zero temperature) the potential reads,
\be 
V(a)\approx \Lambda_{QCD}^4\l 1-\cos\l\frac{a}{F_a}\r\r,
\ee
where $\Lambda_{QCD}$ is the characteristic scale at which the model enters the strong coupling regime. The value of $\Lambda^0_{QCD}$ can be connected to the QCD coupling constant $g_0$ at some reference energy scale (here we take the Planck scale $M_P$). In the SM, the scale $\Lambda^0_{QCD}$ at which the running of $g_0$ brings it to the value $g_0(\Lambda^0_{QCD})\sim 1$ can be found from renormalization group equation \cite{Shifman:1978bx,pdg}, 
\be 
\Lambda^0_{QCD}=M_P\, e^{-\frac{8\pi^2 }{7 g_0^2(M_P)}}\simeq 110 ~\text{MeV}.
\ee
In the inflationary background, the value of $\Lambda_{QCD}$ is modified compared to its usual value $\Lambda^0_{QCD}$. Namely, this scale is defined by the effective QCD coupling $g_{eff}^2=g_0^2/f(h/|X|)$, taken at the reference scale $M_P$. Solving the similar renormalization group equation one obtains,
\be 
\label{LambdaQCD}
\Lambda_{QCD}=M_P\, e^{-\frac{8\pi^2 f(h/|X|)}{7 g_0^2(M_P)}}=M_P \l\frac{\Lambda^0_{QCD}}{M_P}\r^{f(h/|X|)} .
\ee

Therefore, the axion mass is
\be 
\label{m_a}
m_a\sim\frac{\Lambda_{QCD}^2}{F_a}.
\ee

From the lagrangian \eqref{L_scalar_toy} one can obtain the effective axion field normalization during Higgs-dilaton inflation to be 
\be 
F_a^2=\frac{M_P^2}{\xi +\xi_h h^2/|X|^2}\simeq \frac{M_P^2}{\xi}.
\ee

 Considering Higgs-dilaton inflation, in order to obtain the tilt of scalar perturbations to be in agreement with the Planck data, the value of $\xi$ should be smaller than $10^{-2}$ \cite{GarciaBellido:2011de}. Thus, to satisfy the condition $m_a>H$, we need
 
\be 
f(h/|X|)<\frac{\log{H/(2\sqrt{\xi} M_P)}}{\log{\Lambda^0_{QCD}/M_P}}\approx 0.28 .
\ee 
 
This consideration is valid only in the absence of charged under $SU(3)_c$ particles lighter than $\Lambda_{QCD}$, otherwise, the axion mass gets suppressed \cite{diCortona:2015ldu}. Since the SM fermion masses are given by the large value of the Higgs field, they are heavy enough during inflation. However, we need to require that the mass of the additional quark $Q$ is larger than $\Lambda_{QCD}$, to obtain no suppression of the axion mass. In a simple model \eqref{L_Y}, this would provide an unnaturally large value o																								f the coupling $y$. The absence of the large impact to the Higgs mass requires $m_Q<10$ TeV \cite{strumia} at low energies. The described tension is not possible to relax if the value of $y$ is constant. However, we can assume that the value of $y$ depends on the relation $h/|X|$ in such a way that provides with the heavy quark during inflation and light quark in the low energy domain. Clearly, this idea calls for more general study which includes other scale invariant modifications of the lagrangian \eqref{L_scalar_toy}. In the next sections, we construct models which allow for obtaining viable inflation with a sufficiently massive axion.

\subsection{The general scale invariant model}

The most general scale invariant lagrangian for the Higgs field and complex dilaton field coupled to gravity with at most first derivatives reads
\be 
\label{L_scalar}
L_{scalar}=\frac{1}{2}\xi |X|^2 f_2\l\frac{h}{|X|}\r R - \frac{1}{2}f_1\l\frac{h}{|X|}\r|\d_{\mu} X|^2-\frac{1}{2}f_3\l\frac{h}{|X|}\r(\d_{\mu} h)^2- V(h,X),
\ee
with
\be 
V(h,X)=\frac{\lambda}{4}(h^2-\alpha^2|X|^2)^2\l 1+V_1\l\frac{h}{|X|}\r\r.
\ee
Here the functions $f_{1,2,3}$, $V_1$ and the parameter $\alpha$ are defined in such a way that, after the scale symmetry gets spontaneously broken, we recover the SM Higgs potential at low energies. The extra fermions in the model are represented by three sterile neutrinos and the fermion $Q$ charged under $SU(3)_c$ introduced in the previous section. Thus, the lagrangian for the fermionic sector of the model can be written as
\be 
L_{f}=L_{f, \nu MSM} +i\bar{Q}\hat{D}Q+i\bar{\tilde{Q}}\hat{D}\tilde{Q}-y\l\frac{h}{|X|}\r X \tilde{Q} Q + h.c.
\ee 
Here $L_{f, \nu MSM}$ stands for the fermionic part of the $\nu$MSM lagrangian in which the masses of the sterile neutrinos are given by the dilaton vev $\langle X \rangle$ \footnote{It can be done along the lines of \cite{Dias:2014osa} where the Majorana masses of sterile neutrinos come from their Yukava coupling to the Peccei-Quinn scalar.} and $y(h/X)$ is the Yukava coupling which might depend on the values of the Higgs and dilaton fields. As it was discussed earlier, we also modify the QCD lagrangian as in \eqref{QCD}, while the other sectors of the SM remain unaltered. 

\subsection{Einstein frame consideration}

The lagrangian \eqref{L_scalar} is written in the explicitly scale invariant form. However, the presence of non-minimal couplings of scalar fields to gravity makes it not convenient to study the field dynamics during inflation in these variables. This lagrangian can be rewritten in the Einstein frame, in which the action for the gravity takes its canonical form and all scalars are minimally coupled. This can be done performing the Weyl transformation,
\be 
g_{\mu\nu}\rightarrow \hat{g}_{\mu\nu} = \Omega^2 g_{\mu\nu}
  \;~~\text{with}~~
  \Omega^2 = \frac{\xi |X|^2}{M_P^2} f_2\l\frac{h}{|X|}\r,
\ee
and defining the new variables,
\be 
r=|X|,\quad \theta=\frac{h}{|X|}, \quad X=r e^{i a},
\ee
we find that the scalar and gravitational sectors in the Einstein frame, read
\be 
\label{EF1}
\begin{split}
L_{scalar}=\frac{M_P^2}{2} R - \frac{1}{2}\l A(\theta) \frac{(\d_{\mu} r)^2}{r^2} + 2 B(\theta)\frac{\d_{\mu} r}{r}\d^{\mu} \theta +C(\theta)(\d_{\mu}\theta)^2 + F_a(\theta)^2 (\d_{\mu} a)^2\r - \\ -\frac{\lambda \theta^4 M_P^4}{4\xi^2 f_2(\theta)^2}(1+V_1(\theta)).
\end{split}
\ee
Here, the functions $A,~B,~C,~F_a$ are related to the functions in the lagrangian \eqref{L_scalar}, as
\be 
\label{list}
\begin{split}
& A(\theta)=\frac{M_P^2 \left(f_1(\theta)+6 \xi  f_2(\theta)+f_3(\theta) \theta^2 \right)}{\xi  f_2(\theta)},\quad \\ 
& B(\theta)=\frac{M_P^2 \left(3 \xi  f_2'(\theta)  +f_3(\theta) \theta \right)}{\xi f_2(\theta)},\quad \\ 
& C(\theta)=\frac{M_P^2 \left(3 \xi  f_2'(\theta)^2+2 f_2(\theta) f_3(\theta)\right)}{2 \xi  f_2(\theta)^2},\\
& F_a(\theta)^2=\frac{M_P^2 f_1(\theta)}{\xi f_2(\theta)}.
\end{split}
\ee
 
The kinetic term of \eqref{EF1} can be diagonalized by the change of variables
\be 
(r,\theta, a)\rightarrow (\bar{r},\theta, a),\quad \bar{r}=\log{\frac{r}{M_P}}+\int \frac{B(\theta')}{A(\theta')}d\theta'.
\ee
A straightforward calculation reveals that
\be 
\begin{split}
L_{scalar}= \frac{M_P^2}{2} R -\frac{1}{2}\l C(\theta)- \frac{B(\theta)^2}{A(\theta)}\r (\d_{\mu}\theta)^2 + A(\theta)(\d_{\mu}\bar{r})^2+ F_a(\theta)^2(\d_{\mu} a)^2-\\ -\frac{\lambda \theta^4 M_P^4}{4\xi^2 f_2(\theta)^2}(1+V_1(\theta)).
\end{split}
\ee
We can see that the two fields, $\bar{r}$ (hereafter we will refer to it as Einstein frame dilaton --- EF dilaton) and the axion $a$, do not have a (classical) potential. This means that they possess shift symmetry. Therefore, the only field which can be responsible for inflation is $\theta$, whose kinetic term reads
\be
\label{kin}
2 K_{\theta}=\frac{M_P^2}{2\xi f_2^2}\frac{(\d_{\mu} \theta)^2}{f_1+6\xi f_2+f_3\theta^2}\left\lbrace 3\xi f_1 f_2'^2 +f_3\left[2 f_1 f_2 + 3\xi(4 f_2^2-12 f_2' f_2\theta + f_2'^2\theta^2)\right]\right\rbrace.
\ee
Note that the field $\theta$ can be canonically normalized, something that is not possible for the EF dilaton and axion fields, since their kinetic terms depend on $\theta$.

\section{Constraints on the action}

Let us list here several requirements that the lagrangian has to satisfy in order to obtain a viable scenario.
\begin{enumerate}
\item QCD has to be in the strong-coupling phase during inflation which constrains the function $f(\theta)$ in \eqref{QCD}.
\item The axion mass during inflation is constrained from the both sides. The condition $m>H$ provides with zero initial $\theta$-angle and guarantees the absence of isocurvature perturbations. But the axion can not be too heavy, in order to obtain no overproduction of the dark radiation (see Appendix B).
\item We restrict ourselves to inflationary potentials which have only one minimum corresponding to the SM vacuum. Hence we require that at large values of $\theta$, the potential converges to a constant of order $\Delta^2 M_P^2$ with $\Delta=10^{-5}$, to be able to provide the scalar perturbations in agreement with the Planck data \cite{planck}.
\item To obtain the correct values for all the inflationary parameters, we require the following asymptotic form for the inflaton kinetic term at large $\theta$,
\be 
\label{kin_term}
2K_{\theta}=\Lambda^2\frac{(\d_{\mu} \theta)^2}{\theta^2}.
\ee
Although this is not unique choice \footnote{For example, in the usual Higgs dilaton model, the asymptotic form of the inflaton kinetic term reads (in our notifications) $K_{\theta}\propto 1/(\theta^2(\theta^2+\zeta))$ \cite{Karananas:2016kyt}. We leave the study of this case beyond the scope of this paper since it contains more free parameters than the case which we consider here.}, we will use it since it is the simplest one. Note that this form corresponds to a class of $\alpha$-attractor models \cite{Kallosh:2013yoa} which are known to give predictions in a good agreement with the Planck data \cite{planck}.
\item Addressing the question about the quantum stability of the model during inflation, we have to require that the effective suppression scale for all non-renormalizable operators is much larger than the Hubble scale. For the class of models under discussion, we show in Appendix A that this cutoff is nothing else than $\Lambda$. Thus we need $\Lambda \gg H$ \footnote{ Note that the here we discuss the cutoff scale calculated in the inflationary background. The cutoff scale in the vacuum can be smaller than the Hubble scale. In this case, one needs additional assumptions about the UV completion of the model. In the discussed constructions, we postulate the asymptotic shift symmetry for the inflaton field in the Einstein frame for large field values. This assumption protects the flatness of the inflaton potential from possible power law corrections.}.
\item The $SU(3)_c$ charged fermion $Q$ must be heavy ($m_Q>H$) during inflation, otherwise the axion mass is suppressed. On the other hand, a naturally small Higgs mass requires that at low energies its Yukawa coupling $y$ is small enough. This, in turn, corresponds to the mass around $1-10$ TeV. This allows to put constraints on the function $y(\theta)$.

\end{enumerate}

\subsection{General conditions for the functions in the lagrangian}

We start with discussing the conditions that allow for a viable slow roll inflation with a heavy axion. From \eqref{list}, we obtain  
\be 
F_a(\theta)^2=\frac{M_P^2 f_1(\theta)}{\xi f_2(\theta)} .
\ee
For simplicity, let us assume that $F_a(\theta)$ becomes a constant at large $\theta$ and $F_a<H$.  In this case, we have
\be 
\label{cond1}
\frac{f_1(\theta)}{f_2(\theta)}\rightarrow \gamma,~~ \Delta=\frac{H}{M_P}\approx 10^{-5}.
\ee
As for the inflaton kinetic term \eqref{kin}, we can simplify its asymptotic form for large $\theta$ within \eqref{cond1} and assuming $\gamma/\xi\ll 1$. Besides that, let us also assume that the function $f_2$ grows not too fast. Namely, we require $f_2'(\theta)/f_2(\theta)\lesssim c f_2/\theta$ where $c$ is a constant of order unity.\footnote{The equality holds for polynomial functions.} Thus, the kinetic term for $\theta$ can be simplified an is given by
\be 
2 K_{\theta}=\frac{3 M_P^2}{2\theta^2}\frac{(\d_{\mu} \theta)^2}{6\xi f_2/\theta^2 +f_3}\left[\gamma c^2 \frac{f_2}{\theta^2}+f_3 (4-12 c + c^2)\right].
\ee
To start with, let us consider the case in which the terms with $f_2$ dominate both in the numerator and denominator. Under this assumption, we obtain 
\be 
2 K_{\theta}=\frac{M_P^2 c\gamma}{4\xi \theta^2}(\d_{\mu} \theta)^2, ~~\Lambda^2\sim \frac{c \gamma M_P^2}{\xi}\ll M_P^2.
\ee
One can see that this model fails to describe inflation consistently, since its cutoff much smaller than the Planck mass. However, if the terms with $f_3$ dominate, then we get consistent inflation, since
\be 
2 K_{\theta}=\frac{ 3 M_P^2}{2 \theta^2}(\d_{\mu} \theta)^2 (c^2-12c+4), ~~\Lambda\sim M_P.
\ee

To conclude, the conditions for $f_1,~f_2,~f_3,~V_1$ at large $\theta$ leading to the consistent inflationary stage in the presence of a massive axion, read
\be 
\begin{split}
&\frac{f_1(\theta)}{f_2(\theta)}\rightarrow \gamma\ll\xi, \quad \frac{\theta f_2'(\theta)}{f_2(\theta)}\rightarrow c\lesssim 1, \\
&\frac{f_2(\theta)}{\theta^2 f_3(\theta)}\ll 1, \quad V_1(\theta)\rightarrow \Delta^2\frac{\xi^2}{\lambda}\frac{f_2(\theta)^2}{\theta^4}.
\end{split}
\ee
Although this class of solutions might not be unique, it is the one providing with a viable inflationary stage.

Now let us turn to the modification of the kinetic term for the gluons \eqref{QCD}. The function $f$ has to make QCD strongly coupled during inflation, i.e. at large $\theta$. The axion mass provided by the effective strong coupling scale $\Lambda_{QCD}$, reads

\be 
m=\frac{M_P^2}{F_a} \l\frac{\Lambda^0_{QCD}}{M_P}\r^{2 f(h/|X|)} .
\ee
The axion mass must satisfy the inequality
\be
H<m<10^{15}~\text{GeV}\,.
\ee
The upper limit corresponds to the production of the dark radiation, due to the disappearance of the axion mass after inflation. In the next section, this effect will be considered in details.

We focus on the case where the strongly-coupled phase of QCD finishes just after inflation, in order to avoid multiple transitions when crossing the critical value of $\theta$ before reheating. In the next section, we discuss a particular example for the various functions.

Finally, the mass term for the new heavy quark $Q_E$ in the Einstein frame is given by
\be 
L_{Q}=\frac{M_P y(\theta)}{\sqrt{\xi f_2(\theta)}}\tilde{Q}_E Q_E, ~~Q_E=\Omega^{-3/2}Q.
\ee
Here the fermion gets rescaled in order to make its kinetic term canonical. To have a quark with mass 1 TeV in the SM vacuum, we should set 
\be 
y(0)\sim \kappa \sqrt{\xi},~~\kappa=10^{-16}.
\ee
On the other hand, to obtain a heavy quark during inflation, we need
\be 
y(\theta)>\frac{H}{M_P}\sqrt{\xi f_2(\theta)}.
\ee

\subsection{An explicit example}

In this subsection we discuss the set of functions of the simplest form (that is,  polynomials of the lowest degree) which satisfy all aforementioned conditions. The simplest choice of $f_2$ quadratic in $\theta$, which corresponds to the model discussed in Sec. 2.1, does not work. Therefore, we start with the set
\be 
f_1(\theta)=1+a\gamma \theta^4,~~f_2(\theta)=1+\frac{\xi_h}{\xi}\theta^2+a \theta^4,~~f_3(\theta)=1+d\theta^2.
\ee
For this choice, the kinetic term for the inflaton (in the limit $\gamma\ll\xi$) becomes
\be 
2 K_{\theta}=\frac{6 M_P^2(d+4 a\gamma)(\d_{\mu}\theta)^2}{\theta^2(d+6 a \xi)}.
\ee
The required potential correction $V_1(\theta)$ is
\be 
V_1(\theta)=\Delta^2\frac{12 \xi^2 a^2}{\lambda} \theta^4.
\ee
Using the above, we find that the axion constant becomes $F_a=M_P\sqrt{\gamma/\xi}$. The inflaton potential can be expanded as
\be 
V(\theta)=V_0(1+\frac{\mu}{\theta^2}+\dots), ~~\mu=\frac{2\xi_h}{a\xi}.
\ee
The change of variables $\theta=e^{\phi/\Lambda}/\sqrt{\mu}$, with
\be 
\Lambda^2=\frac{6 M_P^2(d+4 a\gamma)}{(d+6 a \xi)}
\ee
makes it possible to canonically normalize the inflaton $\phi$. The potential in terms of $\phi$, reads
\be 
V=V_0(1-e^{-2\phi/\Lambda}+\dots).
\ee
Note that the inflation finishes around $\theta^2=\theta_e^2\sim a\xi/\xi_h$. In order to neglect the next terms in the $1/\theta$ expansions for all functions in the lagrangian until the end of inflation, we must require $\theta_e\gg 1$. To have the cutoff $\Lambda\sim M_P$, it would be enough to set $d\gg a\xi$. The next requirement is to verify that the next terms in the expansion of the kinetic term $K_{\theta}$ can indeed be neglected. After a straightforward calculation, one obtains the extra condition $\xi_h\ll d^2$. Note that,
\be 
d\gg a\xi\gg \xi_h,~~\xi_h\ll d^2,~~\gamma<\xi.
\ee
These conditions do not contradict to each other, so they can be easily satisfied simultaneously. 

Turning to the QCD part of the lagrangian, in order to fulfil the requirements of the previous subsection, we can set \footnote{Notice that the polynomial functions cannot provide with large effective QCD coupling during inflation, since they grow at large $\theta$.}
\be 
f(\theta)=\frac{\theta_e^2+b\theta^2}{\theta_e^2+ \theta^2}. 
\ee
This function interpolates between 1 for small $\theta$ and $b$ for the inflationary region. 
Here we require that the QCD transition happens only once and right after inflation: $\theta_{tr}\sim \theta_e$. The value of $b$ is defined by the desired value of the axion mass $m_a$:
\be 
b=\frac{1}{2}\frac{\log\l\sqrt{\gamma/\xi}\,m_a/M_P\r}{\log\l\Lambda^0_{QCD}/M_P\r}
\ee
For example, if we consider $\gamma=10^{-4}\xi$ than $b$ lies in the range between $0.14$ and $0.16$, in order to provide the axion mass $H<m_a<10^{15}$ GeV.

To satisfy the conditions for the $Q$ quark, we need
\be 
y(\theta)=\kappa\sqrt{\xi}(1+\delta\theta^2),~~\delta>\frac{\Delta\sqrt{a}}{\kappa}\sim 10^{11}\sqrt{a}.
\ee
Note that large values of $\delta$ would not bring the theory to the strong coupling regime because there is an additional suppression of the operators by the cutoff scale. We discuss this in more detail in Appendix A.

 Summarising, we can set for example
\be 
\label{numbers}
a\sim 10^{-4}, ~d\sim 1,~ \delta\sim 10^{10}, ~\xi\sim 10^{3}, ~\xi_{h}\sim 0.01, ~\gamma \sim 10^{-1},~b\sim 0.15.
\ee
These values allow for consistent inflation, and also for an axion and quark $Q$ which are heavy enough during inflation and light at low energies. All additional interactions, in the low energy limit, are suppressed by the scale of the dilaton vev. In Appendix B we prove that the outlined choice of the model parameters is stable with respect to radiative corrections during inflation as well as in the low energy domain. Although it is hard to describe the intermediate stage of preheating in this model, we do not expect any dangerous effects. The rigorous study of reheating is beyond the scope of this paper, we write here some comments on this issue. In order to leave the model in a predictive domain after inflation, the reheating temperature must not be high enough to restore the scale invariance: $T_{reh}\ll M_P/\sqrt{\xi}$. If this condition is satisfied (which should be the case for $\xi\lesssim 10^3$ since we expect $T_{reh}\sim 10^{13}$ GeV, as in the Higgs-dilaton model \cite{GarciaBellido:2011de}), then reheating finally provides the thermalized SM plasma with some admixture of coloured quarks $Q$. The latter are expected to decay via their mixing with the SM quarks. However, we should care about the production of light particles whose coupling to SM is suppressed because it may result in undesired relics in the Universe. We address this issue in the next section.

\section{Phenomenology and predictions}

We start with the discussion of the production of the Goldstone bosons present in the model, i. e. axion and dilaton. Once produced, they remain in the expanding Universe. Their energy density rescales as radiation. These particles do not decay because they have strongly suppressed interactions with the SM particles. For the dilaton production, we expect a significant suppression both for the perturbative and non-perturbative channels, similarly to the conclusions of \cite{GarciaBellido:2012zu}. In this model, we expect the reheating temperature to be comparable to the one in Higgs inflation, $T\sim 10^{13}$ GeV \cite{Bezrukov:2008ut}, or even higher. Due to that, the inflaton has no time to decay into dilatons via the Planck suppressed couplings. 

However, the latter is not true for the axion because, during the fast QCD transition after inflation, it's mass $m_a$ quickly changes from at least $10^{13}$ GeV to zero. This may lead to an efficient production of relativistic axions. In Appendix B, we consider a simple analytical model providing an estimate for the energy density of axions created due to the non-adiabatic vanishing of the mass. If the duration of the transition, $\tau=1/(\beta m_a)$, is smaller than the Hubble time and the period of inflaton oscillations, the axion density is given by
\be 
\rho= 0.003\, m_0^4 \,\log \beta.
\ee
We see that although the actual value of $\beta$ (or $\tau$) is hard to extract from the model, the effect depends on this parameter only logarithmically. Thus, it seems reasonable to have $\log\beta\lesssim 10$. Hereafter we use $\rho \sim m_a^4$ as an order of magnitudes estimate.

Let us compute the amount of axion dark radiation arising from the discussed effect. At reheating, the energy density reads
\be 
\rho_{reh}= \rho\l\frac{a_0}{a_{reh}}\r^4\sim m_0^4 \l \frac{H_{reh}}{H_0}\r^{8/3}.
\ee
Here we used the fact that the preheating stage was matter dominated, as in Higgs inflation \cite{Bezrukov:2008ut}. The transition happened at the moment corresponding to $H_0\sim 0.1 H$ and scale factor $a_0$. The index 'reh' here is related to the moment of reheating. The amount of axion dark radiation measured in the number of extra neutrino species is \cite{GarciaBellido:2012zu}
\be 
\Delta N\approx \frac{3 \rho_{reh}}{\rho_{SM}}, ~~\rho_{SM}=\frac{\pi^2}{30}g T_{reh}^4.
\ee
Plugging in numbers, we obtain
\be 
\Delta N\sim 10^{-2}\l\frac{m_0}{10^{15}~\text{GeV}}\r^4 \l\frac{10^{12}~\text{GeV}}{H_0}\r^{8/3} \l\frac{T_{reh}}{10^{14}~\text{GeV}}\r^{4/3}.
\ee
We see that the amount of the dark radiation crucially depends on the axion mass. It might be significant and available for future CMB observations \cite{Abazajian:2013oma,Errard:2015cxa}. Given the present Planck bound \cite{planck1}, we are not able to increase $m_a$ further than $10^{15}$ GeV. Thus, the axion mass lies between $10^{13}$ and $10^{15}$ GeV.

The discussed model provides a nearly invisible axion at low energies. In this domain, its effective decay constant is $F_a=M_P/\sqrt{\xi}$, which can well be of the order of Planck scale or lower. The axion does not contribute to dark matter anymore. However, as we have seen, it can significantly contribute to dark radiation for some parameter choices. Dark matter can be provided by one of the sterile neutrinos, exactly as in the $\nu$MSM scenario. The other sterile neutrinos can be responsible for the lepton and baryon asymmetry as well as for the SM neutrino masses via the see-saw mechanism.

The possible experimental window for this kind of models may be connected to collider searches of strongly interacting quarks. The LHC limit on new quark masses is $m_Q\gtrsim 1$ TeV \cite{Aad:2015iea}, which is quite close to the naturalness bound $m_Q\lesssim 10$ TeV \cite{strumia}. So the relatively light extra quarks may be probed at LHC and the remaining preferable window for their mass is narrow. 

If the decay constant of the axion is significantly smaller than the Planck scale (which can be the case), then the axion can be probed via its mixing to the photon in a magnetic field. All astrophysical bounds dealing with the axion-photon conversion (cooling of stars \cite{HB}) are relevant for the discussed model as well. Future experiments searching for solar axion emission \cite{IAXO, IAXO1}, and for the light shining through the wall \cite{LSW}, are expected to put stronger constraints on the QCD axion parameters even if it does not contribute to the dark matter. In this case, all experiments specially designed for searches of the dark matter axions would give null results.

\section{Conclusions}

In this paper, we examined the possibility to provide a solution to the strong CP problem in QCD in the framework of spontaneously broken scale invariance. This minimal extension of the Standard Model calls for the complex scalar field (SM singlet) and two Weyl fermions charged under $SU(3)_c$ and $U(1)_Y$ SM groups. The modulus of the scalar field plays the role of the dilaton, while its phase is the QCD axion. Thus, the model (in the SM vacuum) contains two light scalar fields---dilaton and axion---being the Nambu-Goldstone bosons of the broken scale invariance and Peccei-Quinn symmetry, respectively. We show that it is possible to construct a model satisfying all bounds for inflation and QCD axion by the price of considering non-linear lagrangians for the Higgs, gravity and QCD sectors. We prove that, under several conditions for the functions in the lagrangian, one can obtain an effective theory expanded over the classical inflationary background, as well as over the SM vacuum, that is stable with respect to the quantum corrections. Given that, we can conclude that the presented scale invariant models for the QCD axion might be a technically natural, self-consistent effective description for low energy physics as well as for inflation.

The distinct ''signature'' of the discussed scenarios is the presence of a QCD axion whose mass, decay constant and coupling to photons are connected as in the usual KSVZ model \cite{Kim:1979if,Shifman:1979if}. Unlike the usual case which is widely discussed in the literature, this particle does not contribute to dark matter, providing null results for the axion dark matter searches. However, the axion mixes with the photon in the presence of a magnetic field, potentially allowing for its detection in laboratory experiments and astrophysical observations.

The author is indebted to M. Shaposhnikov for bringing the attention to the problem and for the helpful discussions. A.T. thanks also G. Karananas and S. Troitsky for useful conversations. This work was partially supported by the Swiss National Science Foundation. The part of work related to the study of inflation and cosmological consequences of the discussed model was supported by Russian Science Foundation grant 14-22-00161.

\section*{Appendix}

\appendix

\section{Cutoff scale during inflation}

The model described in this paper can be considered as an effective field theory at low energies around the SM vacuum, as well as during inflation. The domain of validity for this description is restricted by the cutoff scale defined as the energy after which the tree level unitarity is violated. Here we perform an estimation of the cutoff scale during inflation in our model in the Einstein frame. We follow the method described in details in \cite{Bezrukov:2010jz}. We expand the lagrangian over the inflationary background and define the cutoff scale as the lowest scale suppressing the non-renormalizable operators in the expansion.

Let us propose that the kinetic term of the field $\theta$ for $\theta\gg 1$ has a form
\be 
2 K_{\theta}=\frac{\Lambda^2}{\theta^2}(\d_{\mu} \theta)^2\l 1+O\l \frac{1}{\theta^2}\r\r.
\ee
Then, the change of variables $\theta=e^{\phi/\Lambda}$ would provide with the canonically normalized scalar field $\phi$. As we have required, during inflation all functions in the lagrangian can be expanded in inverse powers of $\theta$, starting from the constant term.  Under this assumption, for any operator ${\cal O}$ coupled to $\theta$, we can write
\be 
M^l(1+\zeta \theta^{-2}){\cal O}= M^l {\cal O}\l 1+\zeta e^{-2\bar{\phi}/\Lambda} \,\sum_{n=0}^{\infty} \frac{1}{n!}\l-\frac{2\delta \phi}{\Lambda}\r^n\r.
\ee
Here $M^l$ stands for the mass parameter appearing in the lagrangian to compensate the mass dimension of the operator ${\cal O}$. We expand the inflaton field over the inflationary background, $\phi=\bar{\phi}+\delta\phi$. Thus, the cutoff scale $\tilde{\Lambda}$ is the lowest scale which can be obtained from the equation
\be 
\zeta M^l {\cal O}\,  \frac{e^{-2\bar{\phi}/\Lambda}}{n!}\l-\frac{2\delta \phi}{\Lambda}\r^n=\frac{{\cal O}\delta\phi^n}{\tilde{\Lambda}^{n-l}}.
\ee
If $\zeta M^l/\Lambda^l <1$ one can obtain the lowest $\tilde{\Lambda}$ from the term with $n=l+1$,
\be 
\tilde{\Lambda}=\frac{\Lambda}{\zeta} \l\frac{\Lambda}{M}\r^l> \Lambda.
\ee
Therefore, we can take the cutoff scale of the model to be $\tilde{\Lambda}\simeq\Lambda\sim M_P$, given the assumption $\zeta M^l/\Lambda^l <1$ is satisfied (which is usually the case).

Let us check the validity of this conclusion for the mass term of the extra quark $Q$ since we need the large dimensionless constant $\delta\sim 10^{10}$ \eqref{numbers}. The corresponding operator looks,
\be 
L_{Q}=\frac{\kappa(1+\delta\theta^2)}{\sqrt{1+\xi_h\theta^2/\xi+a\theta^4}}M_P \tilde{Q}_E Q_E
\ee
For large $\theta$ it can be expanded as 
\be 
L_{Q}=\frac{\kappa M_P\delta}{\sqrt{a}} \l 1+\l\frac{1}{\delta}-\frac{\xi_h}{2 a\xi}\r\theta^{-2}\r \tilde{Q}_E Q_E.
\ee
We see that the cutoff scale is not lower than $\Lambda\sim M_P$ within the discussed choice of parameters.

Analogically, one can find the cutoff scale for other non-minimal operators in the lagrangian, including $G_{\mu\nu}G^{\mu\nu}$ and the inflaton potential, to be $\Lambda$ during inflation. As for the SM vacuum, all the additional interactions in the model are suppressed by the dilaton vacuum expectation value providing the cutoff scale to be of order $M_P/\sqrt{\xi}$. Therefore, the physics at low energies near the SM vacuum is the same as in the Standard Model.

\section{Radiative corrections to the model parameters}

Let us first discuss the loop corrections that might affect the coupling constants in the low energy domain. All operators suppressed by the large dilaton vev (for example, corrections to the Higgs, dilaton and QCD kinetic terms) actually provide with negligible impact to the running of the dimensionless couplings. This fact, typical for the non-renormalizable models with large cutoff scale, was confirmed for the Higgs inflation \cite{Bezrukov:2010jz} and Higgs-dilaton model \cite{Bezrukov:2012hx}. Thus, we do not expect large loop contributions to the dimensionless couplings from the non-renormalizable operators. From the other side, the small values of these couplings guarantee the stability of the choice of constants \eqref{numbers}. The operators suppressed by the dilaton vev can receive corrections from renormalizable operators. However, these corrections are small provided that all the couplings are weak. 

The separate discussion is needed for the dimensionless couplings $\xi$ and $\xi_h$. The large value $\xi\sim 10^3$ of non-minimal coupling between the dilaton and gravity is stable because the dilaton is very weakly coupled to the SM sector. Similarly to the case of the Higgs inflation, this large value does not lead to significant contributions to other constants, since it translates to the SM sector only through the gravitons. Unlike the dilaton coupling $\xi$, the value of $\xi_h$ cannot be too small for the naturalness reasons. Namely, 
at one loop the SM couplings provide the running of $\xi_h$ \cite{Buchbinder:1992rb,Salvio:2014soa} \footnote{We neglect the contributions from the dilaton which are additionally suppressed by $\alpha^2$.},
\be 
16\pi^2 \frac{d\xi_h}{d \log \mu}=(1+6\xi_h)(2 y_t^2-\frac{3}{4}g_2^2-\frac{3}{20}g_1^2+2\lambda).
\ee
Here $y_t$ is top quark Yukawa coupling and $g_{1,2}$ are the SM gauge couplings. As it follows from this formula, the value $\xi_h \gtrsim 0.01$ is stable under the loop corrections.

Let us now turn to the model expanded over the inflationary background. Technically, it is much simpler to consider the loop corrections in the Einstein frame, in terms of the diagonalized fields. Since the cutoff scale is larger than all the relevant mass parameters of the model (masses of the particles, effective $\Lambda_{QCD}$ scale), we expect all contributions to be suppressed by this cutoff scale. To illustrate this fact we consider the corrections to the kinetic term of gluons. For large $\theta\gg \theta_e$ one can expand the QCD part of the lagrangian,
\be 
\begin{split}
L_{QCD}=&-\frac{1}{4 g_0^2}\l b+2(1-b)\,e^{-2\phi/\Lambda}\r G_{\mu\nu}G^{\mu\nu}= \\& -\frac{1}{4 g_0^2}\l b+2(1-b)\,e^{-2\bar{\phi}/\Lambda}\l 1- 2 \delta \phi/\Lambda+\dots \r\r G_{\mu\nu}G^{\mu\nu}.
\end{split}
\ee
Here we expand the field $\phi$ over its background value $\bar{\phi}$. Before the end of inflation, the last term in the expansion of the function $f$ is small due to the presence of the exponent. One can see that all non-renormalizable interactions are suppressed by the scale $\Lambda$. Their contribution to the value of $b$ is negligible compared to the usual running of the effective QCD coupling $g_{eff}^2=g_0^2/b$ which is already accounted for in the rescaling of the $\Lambda_{QCD}$ \eqref{LambdaQCD}. Therefore, the radiative corrections leave the value of $b$ in the region $0.14<b<0.16$ which is preferred for the axion phenomenology.

The similar expansions over the background inflaton field $\phi$ can be obtained for other non-renormalizable operators. These expansions contain the similar exponential factors $\e^{-\phi/\Lambda}$ which provide with an additional suppression of corrections to the bare parameter values that come from the interactions suppressed by the cutoff scale. Thus, the inflationary observables that depend on the inflaton potential are not significantly affected by the radiative corrections. This fact has been already established for the Higgs inflation \cite{Bezrukov:2010jz}, as well as for the Higgs-dilaton model \cite{Bezrukov:2012hx} in which the inflaton has the exponentially flat potential.

\section{Axion production due to the disappearance of mass}
Let us approximate the axion mass after inflation to be vanishing faster than all cosmologically relevant times (such as the Hubble time and period of the inflaton oscillations). In order to get an analytic solution we use the following mass dependence,
\be 
m^2(t)=m_0^2(1-\tanh(\beta m_0 (t-t_0)), ~~\beta\gg 1.
\ee
For such a function a Bogolubov coefficient and the energy density of axions can be found explicitly \cite{birel},
\be 
\rho=\frac{m_0^4}{2\pi^2}\int\limits_{0}^{\infty}k^3 dk \frac{\sinh^2\l\frac{\pi}{2\beta}(\sqrt{k^2+1}-k)\r}{\sinh\l\frac{\pi k}{\beta}\r \sinh\l\frac{\pi \sqrt{k^2+1}}{\beta}\r} \approx 0.003\, m_0^4 \,\log \beta.
\ee
Note the weak logarithmic dependence of the resulting energy density on the rate of mass vanishing. 

The described effect of axion production seems to be the main impact of the energy density of dark radiation. Other possible sources, such as the Universe expansion and the decay of the inflaton oscillations, can be neglected since they should provide with the result similar to the case of the dilaton. The latter is known to be negligible \cite{GarciaBellido:2012zu}.


\end{document}